# A Pragmatic AI Approach to Creating Artistic Visual Variations by Neural Style Transfer

Chaehan So

*International Design School for Advanced Studies, Design Psychology Lab,*
*Hongik University, Seoul, South Korea*
*cso@hongik.ac.kr*

*On a constant quest for inspiration, designers can become more effective with tools that facilitate their creative process and let them overcome design fixation. This paper explores the practicality of applying neural style transfer as an emerging design tool for generating creative digital content. To this aim, the present work explores a well-documented neural style transfer algorithm (Johnson 2016) in four experiments on four relevant visual parameters: number of iterations, learning rate, total variation, content vs. style weight.*
*The results allow a pragmatic recommendation of parameter configuration (number of iterations: 200 to 300, learning rate: 2e-1 to 4e-1, total variation: 1e-4 to 1e-8, content weights vs. style weights: 50:100 to 200:100) that saves extensive experimentation time and lowers the technical entry barrier.*
*With this rule-of-thumb insight, visual designers can effectively apply deep learning to create artistic visual variations of digital content. This could enable designers to leverage AI for creating design works as state-of-the-art.*

## 1 Introduction

The search for inspiration represents an essential part of visual design. Yet, both design students and professional designers have been observed to consider only a limited range of inspirational approaches (Gonçalves, Cardoso, and Badke-Schaub 2014). One method to overcome this limitation is to consider external stimuli in the idea generation stage. Designers commonly employ this method in their design process, and many studies have confirmed its positive effects. Nevertheless, the exposure to external stimuli also leads to *design fixation*, the narrowing of the creative solution space (Jansson and Smith 1991).

Therefore, it is critical to find the right balance between providing external stimulation on the one side and avoiding design fixation on the other side. To reach this goal, the current work proposes an approach based on artificial intelligence, or AI. The critical factor to avoid design fixation is that the AI element does not generate new solutions at random, but only based on the designer's decision on new style information. Hence, the AI element is not used as a solution generator to replace human creativity but rather as a facilitator for imagining new visual outcomes. This catalyst function of AI may effectively support the creative process by harnessing the latest advancement of artificial intelligence research in the field of deep learning. Particularly interesting for designers and artists, an emerging deep learning topic has become the simulation of human creativity by so-called *generative models*.

## 1.1 Neural Style Transfer

An example of generative models is *neural style transfer*, i.e. the process of merging a visual input with another one. This algorithm gained public attention by an experiment of Hollywood's film-making industry in early 2017: Kirsten Stewart, protagonist of the popular US series "The Twilight Saga", directed the movie "Come Swim" applying neural style transfer and co-authored a research paper that described how this AI algorithm produced the special effects (Joshi, Stewart, and Shapiro 2017).

Technically, this new algorithm built on substantial progress in the development of convolutional neural networks (CNNs) and generative adversarial networks (GANs). Gatys, Ecker, and Bethge (2015) introduced this algorithm to show how the artistic style of painters can be transferred to another image. The neural style algorithm extracts the semantic information of the input image, learns the color and texture information in the style image, and then renders the semantic content of the input image in the color and texture of the style image (Gatys, Ecker, and Bethge 2016).

Designers can create new inspiration by viewing variations that are generated with styles outside their consideration. For example, non-artistic style transfer generates new styles by tiling images of everyday photo motifs like guitars or faces, and succesfully transfers them to the content image (Kenstler 2017). In early conceptual stages, a silhouette can be filled by a style image with neural style transfer that aligns with the silhouette outline (Ramea 2017).

## 1.2 *Visual Applications*

Originally, neural style transfer was demonstrated with common photo motifs like houses or landscapes (Gatys, Ecker, and Bethge 2016). Neural style transfer has later been applied to doodles (Champandard 2016), videos (Huang et al. 2017), artistic improvisation (Choi 2018), and fashion (Jiang and Fu 2017; Zhu et al. 2017).

Recent developments of neural style transfer include using pretrained models for stylization in so-called feedforward networks (Chen and Schmidt 2016). One such approach (Johnson, Alahi, and Fei-Fei 2016) leverages the use of the pretrained models to calculate losses on high-level features (so-called perceptual losses) instead of per-pixel losses.

## 1.3 *Research Question*

Based on the advancement of neural style transfer, it becomes interesting from a design perspective whether it can replace effortful manual design tasks. For example, designers must often create visual variations of a motif like a person portrait, e.g. for album and book covers or posters.

Although the exploration of AI technology looks promising, it is relatively inaccessible to designers. First, it requires the purchase of an advanced graphics card (GPU) that is not standard equipment in desktop or notebook computers. Moreover, a typical AI framework requires the installation of over 50 programs that all must be compatible to each other, including GPU drivers (e.g. CUDA, CUDNN), compilers (e.g. gcc, g++), deep learning libraries (Numpy, Skipy, Scikit-Learn, Hdf5), and loss networks (e.g. VGG-16, ResNet50).

In addition to this technical setup complexity, a designer is confronted with understanding an AI framework that conceptually offers not even remotely a connection with any other design tools or techniques. Therefore, the present work aims at alleviating this accessibility problem for AI to provide a practical guide to the research question:

*How can designers configure a neural style transfer algorithm*
*to produce a range of desirable visual outputs?*

# 2 Method

All results of this paper were generated on an Ubuntu 16.04 LTS virtual machine in the Google Cloud using a Nvidia Tesla P100 GPU with CUDA 9.1 and CUDNN 8.0 libraries.

## *2.1 Technical Framework*

The present study uses the implementation of neural style transfer provided by the Github repository "fast-neural-style" (Johnson 2016). It is implemented in torch (Collobert et al. 2018) and provides an improved version of the neural style transfer algorithm in the Github repository "neural-style" (Johnson 2015). This version represents the original optimization-based algorithm introduced by Gatys, Ecker and Bethge (2015).

The optimization-based algorithm is provided by the script "slow_neural_style.lua" and allows many configuration options. With direct consequence on processing speed, one can configure the options num_iterations (number of iterations processed), save_every (image generation after save_every iterations), GPU (use GPU or CPU). Obviously, the highest impact on processing time is determined by the processing unit: Image processing becomes impractical in CPU mode even if 8 or 16 CPUs are used. GPUs offer processing speeds by a factor of 10 to 100 faster than CPUs.

To modify the stylization outcome, one can configure many detailed options of the loss network. The interface allows to determine the content layers, style layers, style target (gram matrices vs. spatial average) and the choice of the optimizer (commonly LBFG-S or Adam algorithm). All these options require a lot of expertise to understand the direction and magnitude of parameter changes. Configuration parameters accessible to non-experts include the learning rate and the ratio between content weights and style weights.

## *2.1.1 Input Images*

The input images were chosen from celebrity portraits because they allow a relatively easy evaluation of the degree to which neural style transfer maintains the human quality in a recognizable pattern. Two male portraits (Charlie Puth, Will Smith) and two female portraits (Jessica Alba, Ellie Goulding) were selected as input.

Preliminary experiments provided the insight that pictures with high variance in the background were evaluated as part of the foreground and thus stylized the same way as the foreground. To avoid this effect, portraits were only selected if they showed a clear separation to the background.

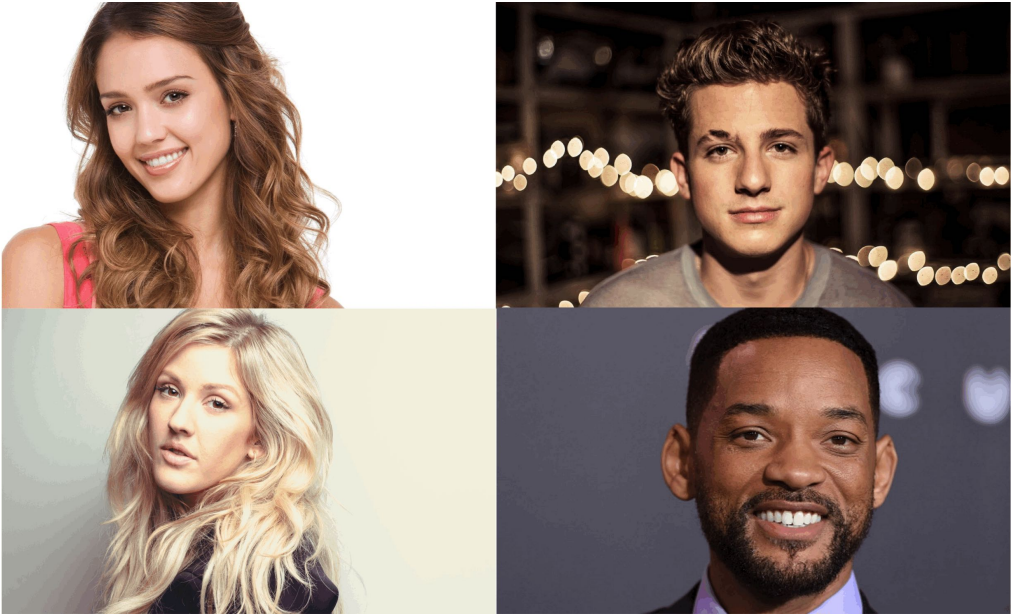

Figure 1: Input Images

### 2.1.2 *Style Images*

The style images were chosen from three different categories: black & white patterns, cloth design, and abstract arts.

The algorithm does not extract the style image's structural information but its color and texture information. This is to ensure that the input image's structural information (like face and body line structure) can be recognized in the output information. In contrast, style images have the most impact on the output image if they contain clear texture information of finer granularity that separates clearly from the background.

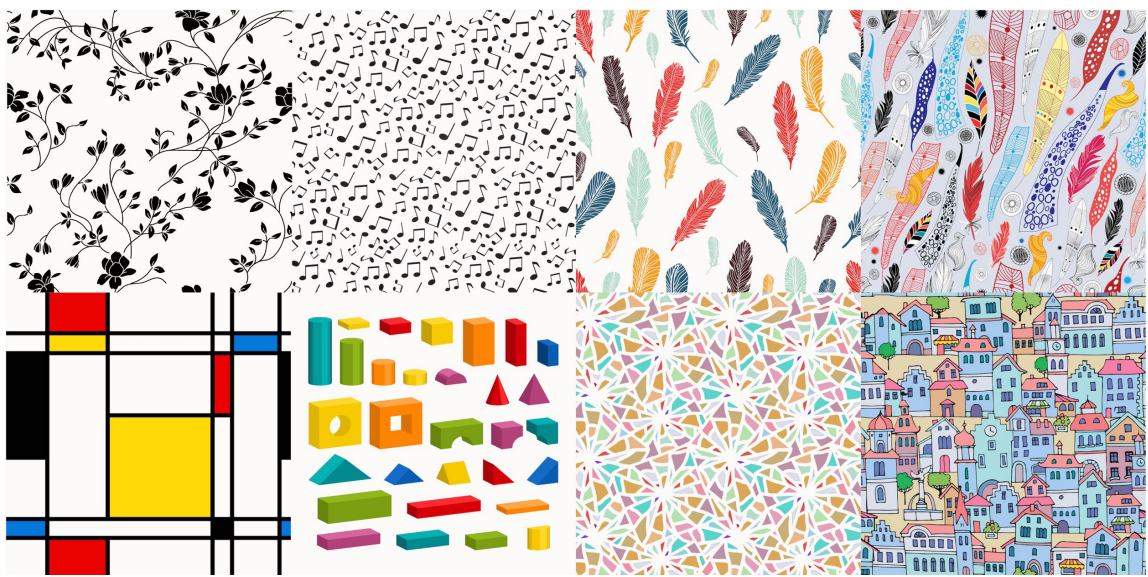

Figure 2: Style Images

*2.1.3   Qualitative Evaluation*

To evaluate the quality of stylization, the outcome images were evaluated according to the best trade-off between application of color and texture from the style image and recognizability of the human impression. Ideally, the resulting image should enable to recognize the person but stylize it to a degree that is perceived as artistic.

*2.1.4   Experiment 1: Number of Iterations*

The neural transfer algorithm by Gatys, Ecker and Bethge (2015) applies the L-BFGS optimizer on forward and backward iterations through the VGG-16 loss network. Johnson, Alahi and Fei-Fei (2016) found that the optimization is successful within 500 iterations in most cases. To save processing time, the first experiment aims to find a lower value for iterations that still renders satisfactory results. Therefore, it compares visual results for 100, 200, 300, 400, and 500 iterations.

*2.1.5   Experiment 2: Learning Rate*

The framework allows to change the optimizer from LBFGS to Adam. The Adam optimizer allows to set different values of the so-called learning rate. This parameter specifies the step size in which weights are updated during the optimization. The default learning rate is set to 1e-3. Therefore, the second experiment explores the impact of learning rates for the Adam optimizer for 1e-0, 5e-0, 1e1, 2e1, 4e1, 6e1.

*2.1.6   Experiment 3: Total Variation*

The regularization on the transformation network's output shows a total variation that can be set by the parameter tv_strength. Its default value is set at 1e-6. The higher the total variation is set, the smoother the output is rendered spatially. Therefore, the third experiment investigates variation values that are higher than the default setting, in particular 1e-8, 1e-6, 1e-4, 1e-2, 1e-1, 1e-0.

*2.1.7   Experiment 4: Content to Style Weight*

The parameter content_weight is set as a relative value in relation to style_weight. This content to style weight ratio defines the degree of importance given to the input image vs. style image for rendering the output image. The default setting is 1:1 or 100:100. Therefore, the fourth experiment explores results for the content to style weight rations 10:100, 50:100, 100:100, 200:100, and 300:100.

# 3 Results

## 3.1 Experiment 1: Number of Iterations

Stylizing the portrait of Charlie Puth reveals that the output images converges between 200 and 300 iterations for most styles to a satisfactory result (see Figure 3). In other words, although more iterations render finer structure details from the input image, more than 300 iterations don't seem to be necessary except on some rare occasions. The biggest noticeable improvement happens between 100 and 200 iterations, indicated by the highest drop in the loss function.

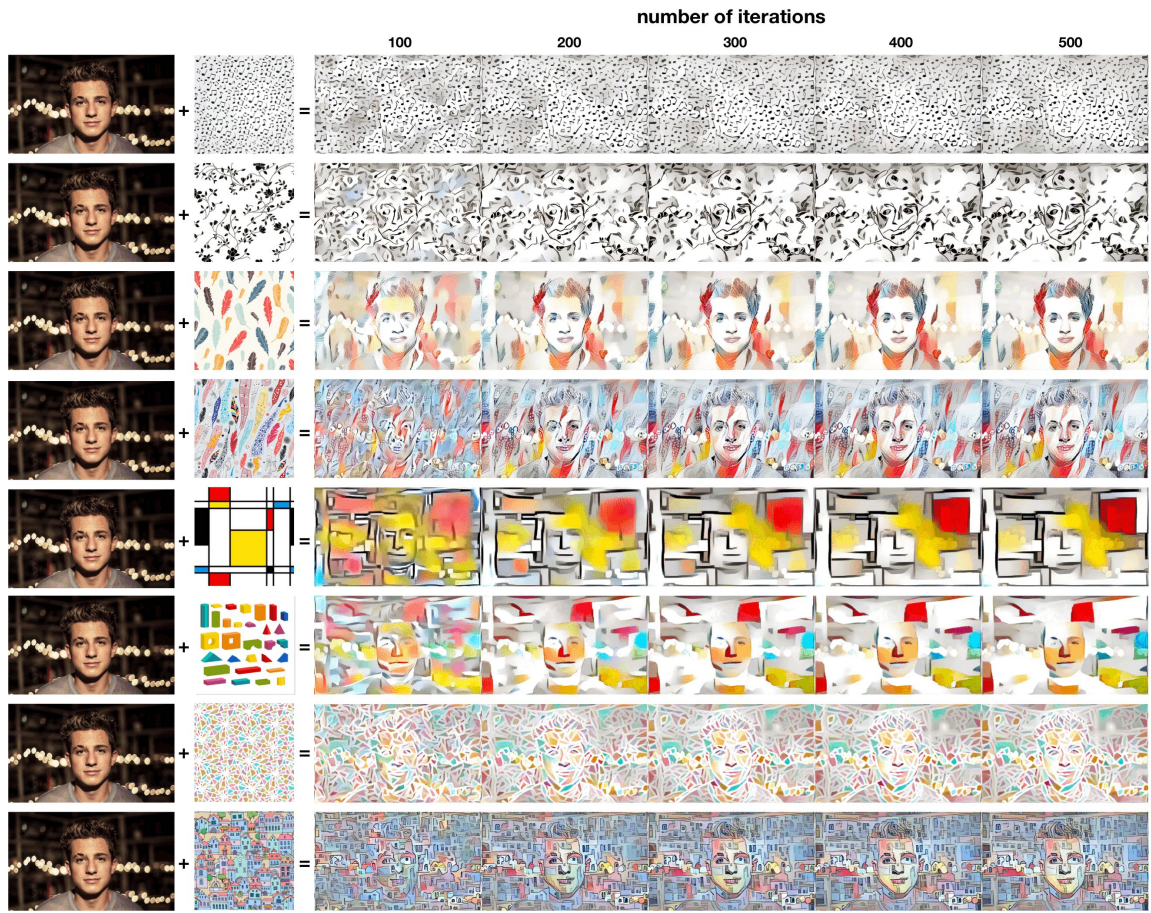

Figure 3: Experiment 1 - Number of Iterations

## 3.2 Experiment 2: Learning Rate

When using Adam as optimizer instead of the default LBFG-S optimizer, the learning rate parameter can be specified. The learning rate represents the step size or incremental change for each iteration when minimizing a loss function. As Ruder (2016) points out, a too small learning rate slows down the convergence to the minimum, whereas a too high learning rate can cause fluctuation around the minimum and thus hinder convergence. Stylizing Jessica Alba (see Figure 4) reveals that the default learning rate of 1.0 (1e-0) is too small to render any portrait hint, whereas 5.0 (5e-0) starts to render face structures in rare cases, and 10 (1e1) provides better results in some cases. Yet, the sweet spot lies at higher learning rates between 2e1 and 4e1 – these rates render recognizable faces with style images that contain visual elements of finer granularity (e.g. feathers or mosaics). However, insufficient results are obtained if style images have finer texture but little texture variance (e.g. musical notes).

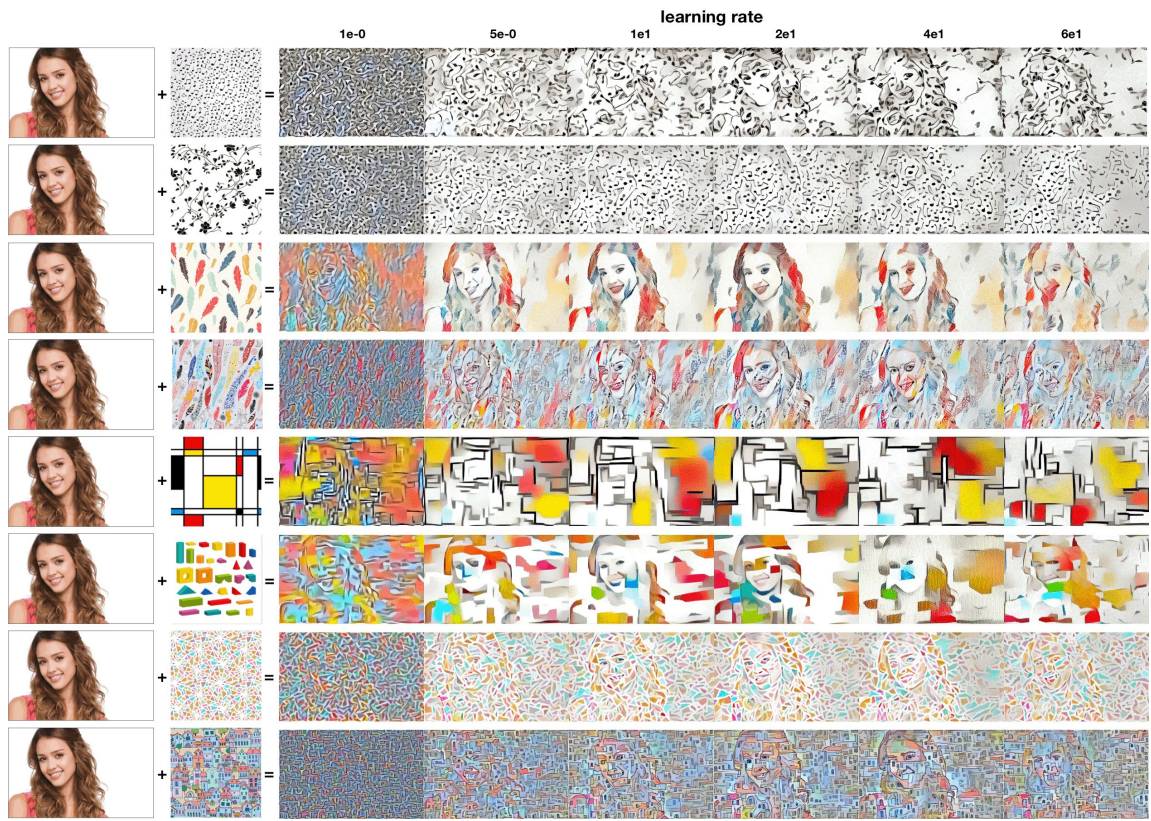

Figure 4: Experiment 2 - Learning Rate

## 3.3 Experiment 3: Total Variation

The total variation is supposed to control the smoothness of rendering. Looking at the visual results of stylizing Will Smith (see Figure 5), the default total variation of 1e-6 turns out to be the best value to render recognizable faces regardless of the style image. Smaller values (1e-8) and higher values (1e-4) render similar and sometimes slightly better stylizations depending on subjective preference.

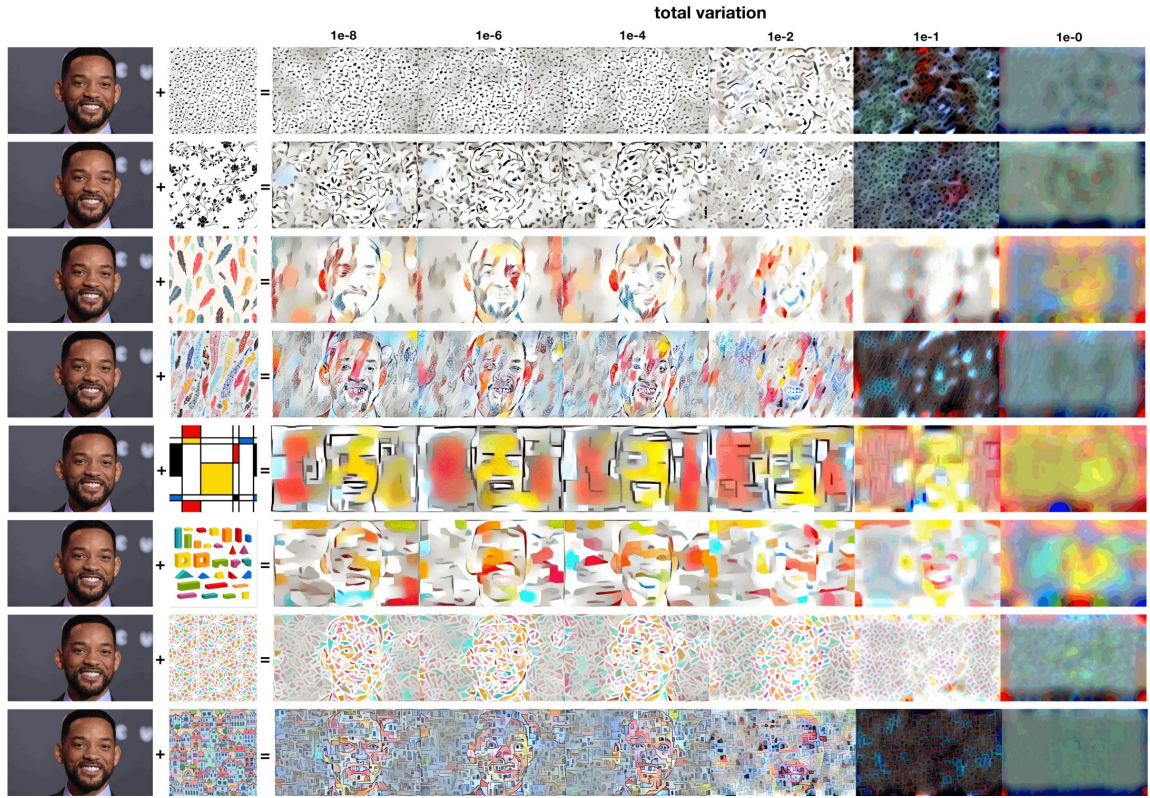

Figure 5: Experiment 3 - Total Variation

## 3.4 Experiment 4: Content to Style Weight

The content and style weight determines the degree of importance for the input image vs. the style image for rendering the output image. Although the two parameters can be configured separately, they only make sense together in their ratio. The default ratio is 1:5, i.e. the style weights are five times larger than the content weights.

The ideal configuration for an artistic result is a trade-off between two optimization challenges:

On the hand, it is important to clearly let the eye recognize the person while not coming too close to a realistic depiction (too much content weight). The results (Figure 6) show that the *maximum* configuration for this goal is 200:100. On the other hand, the output image should reveal the distinct artistic style (color and texture) while not coming too close to the original style image (too much style weight). The results show that the *minimum* configuration for this goal is 50:100.

Despite these recommendations, which content to style weight ratio is chosen depends largely on a conscious decision about the expected outcome. For example, although the 10:100 content to style weight ratio does not allow to recognize the face with most of the style images, this depicted distortion could still be a desired effect.

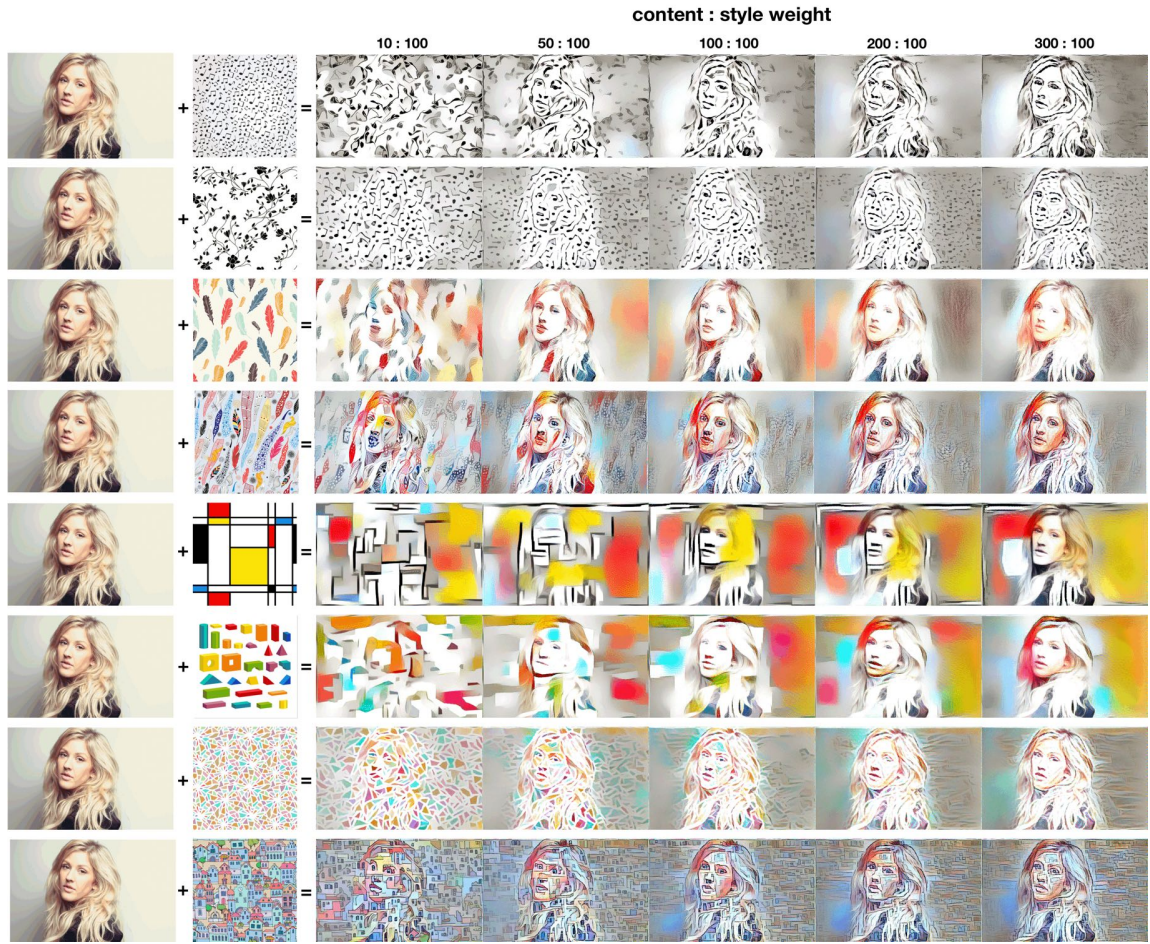

Figure 6: Experiment 4 - Content vs. Style Weight

# 4 Conclusion

The four experiments allow a relative clear recommendation for configuring the neural style framework by Johnson (2016): Best results can be expected with number of iterations between 200 to 300, learning rate between 2e-1 and 4e-1, total variation between 1e-4 and 1e-8, and content weights vs. style weights between 50:100 to 200:100. Stylizing portrait images with this configuration lets the eye perceive an artistic visualization while still recognizing the person portrayed. Designers can thus use neural style transfer as an effective design tool for creating artistic visual variations.

Designers are encouraged to try out the presented framework by Johnson (2016) because its Github repository is one of the best documented implementations of neural style transfer. Acquiring such experience with this algorithm will be extremely beneficial to designers because they can leverage it with further advancement in this AI field – which ultimately will enable them to create design-driven state-of-the-art.